\documentclass[prl,twocolumn]{revtex4-1}

\usepackage{graphicx}
\usepackage{amsmath}
 \usepackage{amssymb}
\usepackage{amsfonts}
\usepackage{comment}
\usepackage{color}
\usepackage{hyperref}
\usepackage[makeroom]{cancel}
\usepackage[normalem]{ulem}

\begin{document}


\title{Interaction-induced transparency for strong-coupling polaritons}

\author{Johannes Lang$^{1}$}
\author{Darrick E. Chang$^{2,3}$}
 \author{Francesco Piazza$^{4}$}

\affiliation{$^{1}$\small Physik Department, Technische Universit\"at M\"unchen, 85747 Garching, Germany}
\affiliation{$^{2}$\small ICFO-Institut de Ciencies Fotoniques, The Barcelona Institute of Science and Technology, 08860 Castelldefels (Barcelona), Spain}
\affiliation{$^{3}$\small ICREA-Instituci\'o Catalana de Recerca i Estudis Avan\c{c}ats, 08015 Barcelona, Spain}
\affiliation{$^{4}$\small Max-Planck-Institut f\"ur Physik komplexer Systeme, 01187 Dresden, Germany}

\begin{abstract}
The propagation of light in strongly coupled atomic media takes place
  through the formation of polaritons - hybrid
quasi-particles resulting from a superposition of an atomic and a photonic excitation.
Here we consider the propagation under the condition of electromagnetically-induced
transparency and show that a novel many-body
  phenomenon can appear due to strong,
dissipative interactions between the polaritons. 
Upon increasing the photon-pump strength,
we find a first-order transition
between an opaque phase with strongly broadened polaritons and a
transparent phase where a long-lived polariton branch with highly tunable
occupation emerges. Across this non-equilibrium phase
  transition, the transparency window is reconstructed via nonlinear
interference effects induced by the dissipative polariton interactions.
Our predictions are based on a systematic diagrammatic expansion
  of the non-equilibrium Dyson equations which is quantitatively valid,
  even in the non-perturbative regime of large single-atom
  cooperativities, provided the polariton interactions are
  sufficiently long ranged. Such a regime can be reached in
photonic crystal waveguides thanks
to the
tunability of interactions, allowing to observe the interaction-induced-transparency transition even at low polariton densities.
\end{abstract}

\maketitle

\emph{Introduction.---}
Slow light in coherent media via electromagnetically induced
transparency (EIT) \cite{Fleischhauer2005}  has become a cornerstone
of quantum optics, allowing for instance to store and even shape
individual photons \cite{Fleischhauer2002, Vernaz2018, Pursley2018}. EIT is a linear optical
phenomenon resulting from destructive interference between dipole-excitation pathways. Combined with strong atomic Rydberg interactions, EIT has also turned out to be a key ingredient in achieving
single-photon non-linearities \cite{chang_vuletic_lukin_2014,firstenberg_review_2016},
potentially allowing for efficient transmission, manipulation, and storage of quantum
information \cite{quantum_internet}. 
On the other hand, single-photon
nonlinearities pave the way for the study of novel quantum many-body
phenomena with strongly interacting photons
\cite{chang_vuletic_lukin_2014}.

In this letter, we consider EIT polaritons, mixing
a set of electromagnetic modes with internal atomic excitations, in the presence of strong,
partially dissipative interactions. 
The latter typically destroy the EIT window \cite{firstenberg_review_2016, bajcsy_2009}.
However, under certain conditions, at a threshold value of the rate at which
photons are injected into the system, we find a first-order phase
transition in the driven/dissipative steady-state, separating an opaque
phase (OP) with a low density of strongly broadened polaritons from a transparent
phase (TP),
characterized by the existence of a long-lived polariton branch with a
high spectral density. The novelty and peculiarity of this transition
resides in the fact that in the TP the transparency window is
reconstructed via nonlinear
interference effects induced by the dissipative interactions, in a
process that we name interaction-induced transparency (IIT). The
first-order transition is accompanied by a bistable region where the
OP and TP coexist. 

\begin{figure}[t]
\begin{center}
\includegraphics[width=\columnwidth]{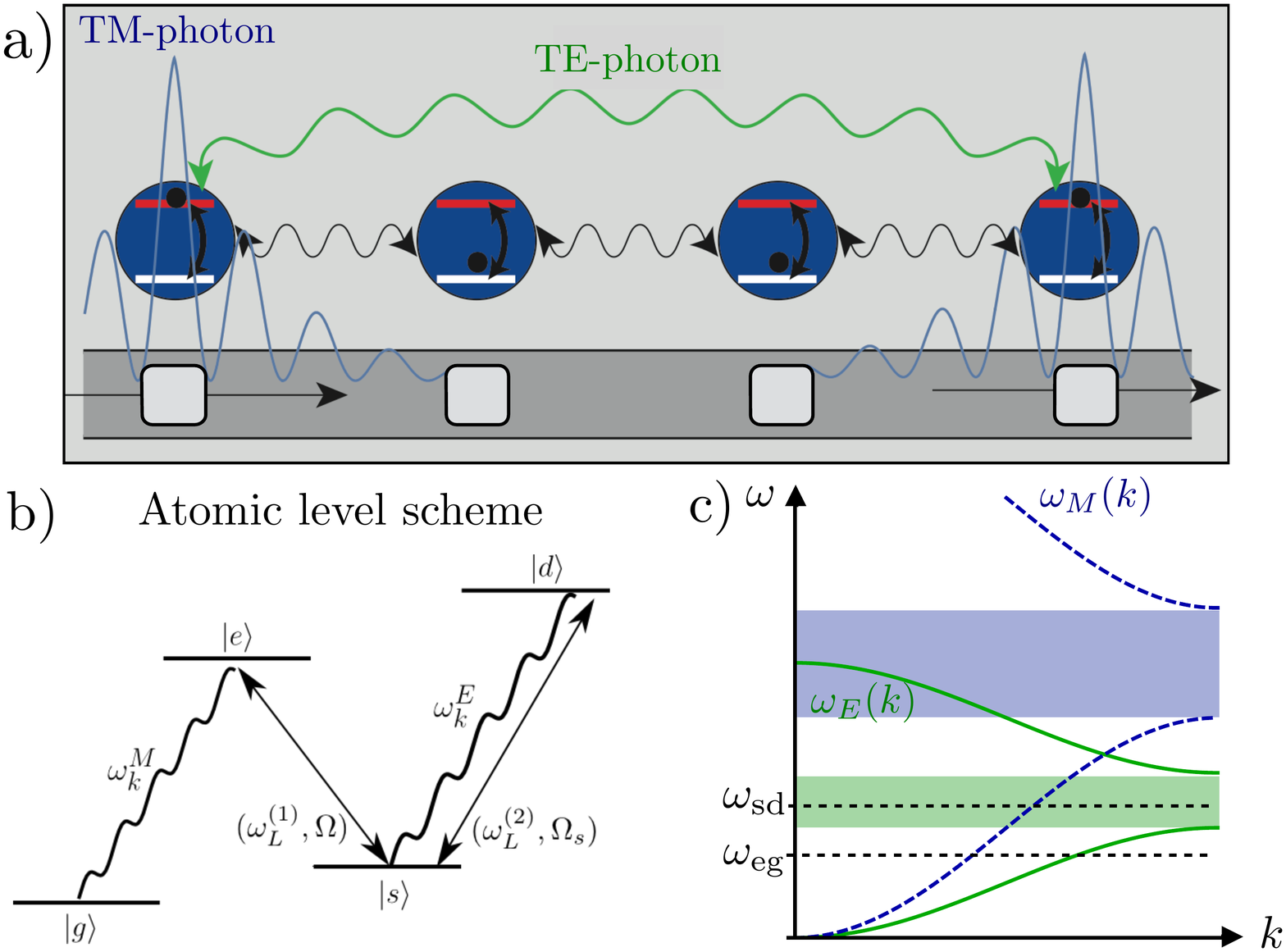}
\caption{a) Proposed realization of IIT: An chain of atoms are fixed
  in a periodic arrangement in the evanescent field of the PCW. The atoms
  interact via internal electronic transitions (shown in b)) with
two continua of photonic modes of the PCW with different transverse
polarization. Panel c) shows where the atomic transitions lie relative
to the band structure of both polarizations.}
\label{fig:setup}
\end{center}
\end{figure}

We employ a diagrammatic expansion of
the Dyson equations for the non-equilibrium response and correlation
functions. This novel approach, described in detail elsewhere
\cite{lang_EIT_long}, allows for quantitative predictions even in the
non-perturbative regime of large single-atom cooperativities (where
IIT takes place), provided the polariton interactions are sufficiently long ranged.

We consider in
  detail a possible implementation of IIT with photonic crystal waveguides
(PCWs). There, the polaritonic excitations can be made to
strongly interact via localized, non-propagating photons within a bandgap (see also Fig.~\ref{fig:setup}).
The engineered photon band-structure potentially allows to control not only
their dispersion but also both the strength and the range of
interactions \cite{Douglas2015a,douglas_molecules_2016,shi2015multiphoton}, as well
as the coupling with the environment \cite{asenjo_radiance_2017}.


\emph{Model and approach.---}
We consider an array of four-level atoms as illustrated in Fig.~\ref{fig:setup}(b). 
The $|g\rangle-|e\rangle$ transition between the atomic ground and excited states is
coupled to guided photons (e.g. a Bloch band of transverse magnetic
polarization in a photonic crystal waveguide) with dispersion
relation $\omega_M(k)$, while $|e\rangle$ is coupled to an additional
metastable state $|s\rangle$ via a coherent control field with Rabi
frequency $\Omega$. In such a configuration the propagating photons, that are near-resonant with the $|g\rangle-|e\rangle$ transition, hybridize
with atomic excitations to $|s\rangle$ which an lead to EIT \cite{Fleischhauer2002}. In particular, the destructive interference between the $|g\rangle\to|e\rangle$ and $|g\rangle\to|e\rangle \to|s\rangle \to|e\rangle$
excitation pathways suppresses population of $|e\rangle$ and the associated spontaneous emission and dissipation of
the propagating photons within a narrow frequency and momentum
window. The polaritons associated with this otherwise linear optical
effect can be made to interact strongly by coupling $|s\rangle$ to a
second excited state $|d\rangle$ by means of a second control field of Rabi
frequency $\Omega_s$. We assume that the $|s\rangle-|d\rangle$
transition is slightly off-resonant with respect to a separate photonic band
with dispersion relation $\omega_E(k)$. As illustrated in
Fig.~\ref{fig:setup} in a PCW the transition can be engineered to lie
within the bandgap of e.g. the transverse electrically polarized
photons or, alternatively, below the mass gap of photons in tapered
fibers. An atomic excitation $|d\rangle$ is then unable to
spontaneously emit a guided photon due to the absence of resonant
modes. Instead, a photon becomes ``bound'' around the atomic
excitation, which facilitates strong interactions with nearby atoms \cite{kimble_2014_crystal}.
Due to the finite life times of both state $|d\rangle$ and the $E$ photon, these are in fact mostly dissipative. Furthermore the atoms
are assumed to be fixed in a periodic arrangement, which can be
achieved using tweezers \cite{Thompson1202, Endresaah3752}  or the evanescent field of PCWs \cite{chang_njp_photcrys_2013}.
We will consider the linear regime of small atomic excitation
densities, where we can replace spin operators with bosonic
creation/annihilation operators
$
\hat{\sigma}_{ee}\to \hat{a}^\dagger_e\hat{a}_e,
\hat{\sigma}_{eg}\to \hat{a}^\dagger_e\hat{a}_g,
$
and similarly for the other atomic transitions.
The Hamiltonian of the system is given by a free part (setting
$\hbar=1$) 
\begin{align*}
\hat{H_0}=\sum_z\!\sum_{j=e,s,d}\!\!\omega_j \hat{a}^\dagger_j(z)\hat{a}_j(z)+\int_k\sum_{j=M,E}\!\!\omega_j(k)\hat{a}_j^\dagger(k)\hat{a}_j(k)
\end{align*}
with dispersions $\omega_E(k)\simeq\omega_E(k_0)+\alpha_E(k-k_{0})^2$
and $\omega_M(k)$, where the precise form of the latter depends on the
physical realization and is of no qualitative relevance for the
following results. Additionally, the atoms interact with laser and guided photons via
\begin{align}
\label{eq:ham_initial}
&\hat{H}_{\rm int}=\sum_z\bigg[\Omega e^{-i \omega_L^{(1)}t}\hat{a}^\dagger_e(z) \hat{a}_s(z) +\Omega_s e^{-i\omega_L^{(2)}t}\hat{a}^\dagger_d(z) \hat{a}_s(z) \nonumber\\
&+\!\!\!\!\sum_{j=M,E}\int_kg_j\hat{a}_j(k) e^{i k z} u_k^j(z) \hat{a}^\dagger_{\mathrm{exc}(j)}(z) \hat{a}_{\mathrm{gs}(j)}(z) \!+\! h.c.\bigg]
\end{align}
with the notation $\mathrm{gs}(M/E)=g/s,\mathrm{exc}(M/E)=e/d$, and
where $\sum_z$ runs over the sites of the atomic lattice and $u_k^{M/E}(z)$ represents the periodic part of the
Bloch functions of either polarization at quasi-momentum $k$. We use
the standard convention $\sum_z e^{ikz}=2\pi\delta(k)$ with lattice
constant $a=1$ and $\int_k=\int_{-\pi}^\pi\frac{dk}{2\pi}$ for the
integration over quasi-momenta. 
The incoherent dynamics is described by 
\begin{align}
\!\!\big(\!\!\!\sum_{\scriptsize\begin{array}{c}j\!=\!e,d,z\end{array}}\!\!\!\!\gamma_j\mathcal{D}[\hat{a}_j(z)]\!+\!L\!\int_k\!(\!\kappa_s\mathcal{P}[\hat{a}_M(k)]+\!\!\!\!\sum_{\; j=M,E}\!\!\!\!\kappa_j\mathcal{D}[\hat{a}_j(k)])\big)\hat{\rho}\nonumber
\end{align}
with the dissipator 
$
\mathcal{D}[\hat{L}]\hat{\rho}=-(\{\hat{L}^\dagger\hat{L},\hat{\rho}\}-2\hat{L}\hat{\rho}\hat{L}^\dagger)/2
$
describing the spontaneous decay of the excited atomic states
plus the photon losses out of the guided modes, and the pump
$\mathcal{P}[\hat{L}]\hat{\rho}=-(\{ \hat{L}\hat{L}^\dagger+\hat{L}^\dagger \hat{L},\hat{\rho}\}-2\hat{L}^\dagger\hat{\rho}\hat{L}-2\hat{L}\hat{\rho}\hat{L}^\dagger)/2
$
homogeneously injecting M-photons into the waveguide. 
The incoherent pump results from coupling with a reservoir
and therefore cannot generate an inversion in the $k=0$ mode, which
excludes the phenomenon of polariton condensation \cite{carusotto_rev_2013,sieberer_2014}.
\begin{figure}[t]
\begin{center}
\includegraphics[width=\columnwidth]{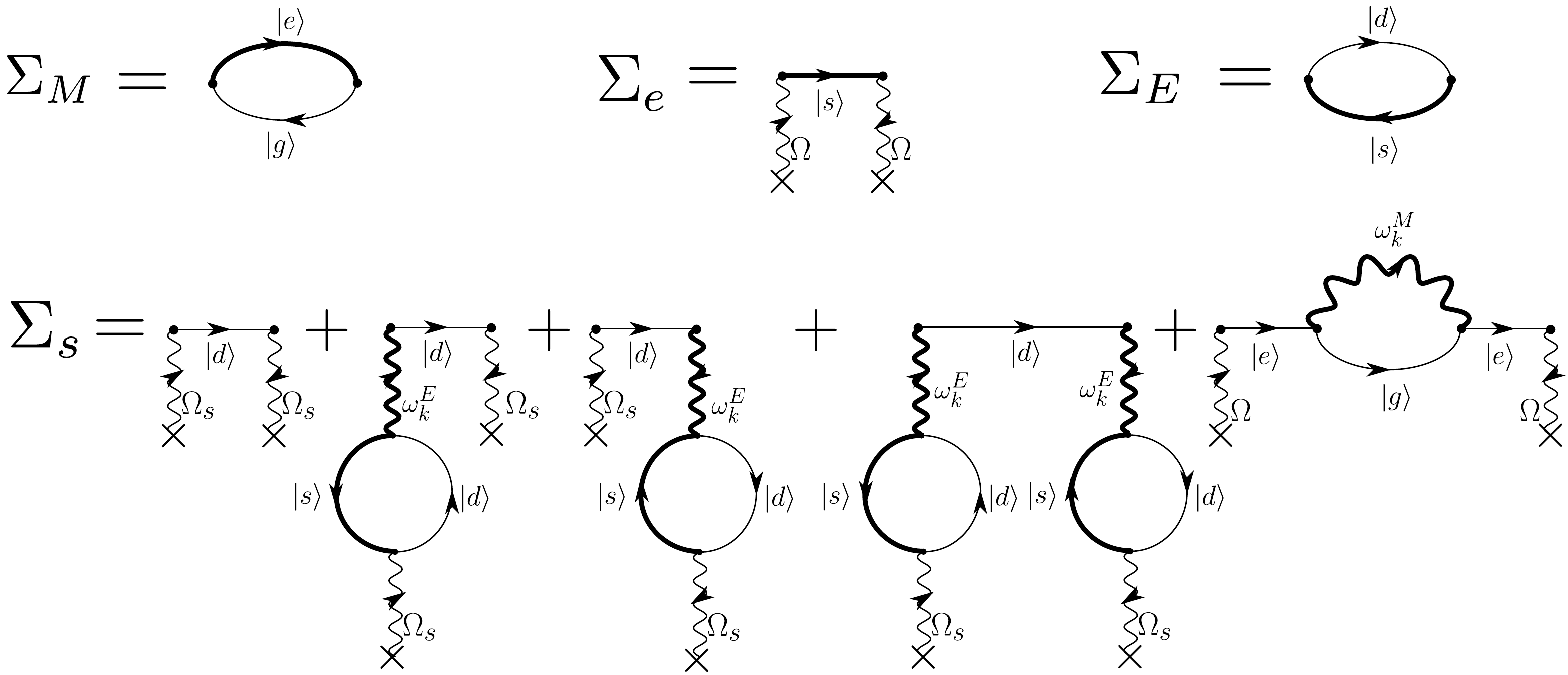}
\caption{Self-consistent Hartree diagrammatic approach to the
  computation of self-energies for the magnetically/electrically
  polarized photons $\Sigma_{M/E}$ and the atoms in the metastable
  state $\Sigma_s$. Internal solid/wiggled lines represent atom/photon
  propagators while wiggled legs ending in a cross indicate insertions
  of a static laser field. Bold lines correspond to propagators
  including self-energy corrections, which implies a self-consistent
  treatment. In order to account for the fact that an atom can be
  excited only once, all diagrams where the same
  atoms appears more than once in the same state must be removed \cite{lang_EIT_long}.}
\label{fig:N_scheme_diagrams}
\end{center}
\end{figure}

Since we are interested in the steady state of this quantum driven/dissipative
interacting system in the thermodynamic limit, we employ the
Keldysh non-equilibrium functional integral formalism \cite{kamenev_book,sieberer_Keldysh_review,Lang2016} to compute
single-particle Green's functions (or propagators). 
Due to the absence of thermal equilibrium, 
there are in principle two independent propagators, the retarded
$
i\left(G_j^R(x,x')\right)=\theta(t-t')\langle [ \hat{a}_j(x),\hat{a}_j^{\dag}(x)]\rangle
$,
and the Keldysh Green's function
$
i\left(G_j^K(x,x')\right)=\langle \{ \hat{a}_j(x),\hat{a}_{j^\dag}(x)\}\rangle
$
with $j=g,e,s,d,E,M$ labeling the degree of freedom and
$x=z,t$ being the space-time coordinate.
These two propagators satisfy two coupled non-equilibrium Dyson
equations, for which we
developed a controlled diagrammatic treatment of the interaction
processes that is described in detail elsewhere
\cite{lang_EIT_long}. 

Here we limit ourselves to introducing the 
self-consistent Hartree (SCH) theory illustrated diagrammatically in
Fig.~\ref{fig:N_scheme_diagrams} and formally in the Supplemental
Material. This SCH theory is the simplest set of diagrams giving rise
to IIT, allowing for the clearest illustration of the
  phenomenon. As shown in \cite{lang_EIT_long}, the SCH approach can
  be extended in a controlled manner (by including a few more diagrams) to
  become quantitative even in the non-perturbative regime of large
  single-atom cooperativities considered below.
The $M$-photon
self-energy $\Sigma_M^R$ of Fig.~\ref{fig:N_scheme_diagrams}, is the susceptibility (or polarization function) describing the
modified propagation in the atomic medium. The contribution to
  $\Sigma_M$ involving a bare $e$-propagator adds up with the one
  involving the insertion of $\Sigma_e$, which corresponds to the interference of pathways underlying EIT. We include the
modified propagation due to the medium also for the E-photons
through the self-energy $\Sigma_E$. Finally, the interactions between
$s$-atoms mediated by E-photons are taken into account at the
Hartree level by the first four diagrams in $\Sigma_s$. Since the
level $s$ is not driven, those diagrams vanish unless we dress the $s$
propagator with the last diagram, which creates a finite occupation in
$|s\rangle$ with the help of an M-photon. In a diagrammatic loop expansion, this means that the first non-vanishing
correction to the EIT propagation of the M-photons appears at the
3-loop level. Our approach is non-perturbative in two ways:
i) the Dyson equation implies a resummation of an infinite number
of self-energy insertions, which is for instance needed to 
describe EIT; ii) the self-energies
contain bold $s,M$ and $E$ propagators, indicating a self-consistent
treatment. In particular, this allows to account for the fact that
the E-photon-mediated interactions are screened due to polarization
effects and that the population transfer into the interacting
$|s\rangle$ state via
the M-photons takes place only within the EIT transparency window.

The SCH approach of Fig.~\ref{fig:N_scheme_diagrams} captures the
energy-shift and modified damping of the atoms in state $|s\rangle$ due to the
E-photon mediated interactions, but neglects the scattering
processes involving energy and momentum transfer. 

As explained in \cite{lang_EIT_long}, this approach is justified in
the regime of a small single-atom cooperativity
$C_{E}=g_{E}^2/(\kappa_{E}\gamma_{d}L_{E})\ll 1$,
where $L_{E}$ is the effective (i.e. including
polarization effects) propagation range of the E-photons in units
of the atom spacing $a$. 
$L_E$ also corresponds to the effective interaction range between the
atoms in state $|s\rangle$, which when $\kappa_E\gg\omega_L^{(2)}-\omega_E(0)$ simply
reads $L_E\simeq \sqrt{\alpha_E/\kappa_E}$. On the other hand, inside
the transparency window M-photons propagate essentially freely, so that
$L_M\simeq v_M/\kappa_M$, where we assumed a linear dispersion
with slope $v_M$ inside the EIT window (see
Fig.~\ref{fig:freqmom_res_corr}). Note that our approach does not require a small single-atom cooperativity $C_{M}=g_{M}^2/(\kappa_{M}\gamma_{e}L_{M})$. Therefore, the strong coupling regime, where
\begin{align}\label{eq:strong_coupling}
p L_E C_E C_M \frac{\Omega_s^2}{\Omega^2}\gg 1,
\end{align}
with the pump ratio $p=\kappa_s/\kappa_M$ fixing the density of
excitations, can be described faithfully. 
This is necessary for interaction-induced shifts to become important, so that non-perturbative
effects like IIT can appear \cite{lang_EIT_long}. 

\emph{Results.---}
\begin{figure}[t]
\includegraphics[width=\columnwidth]{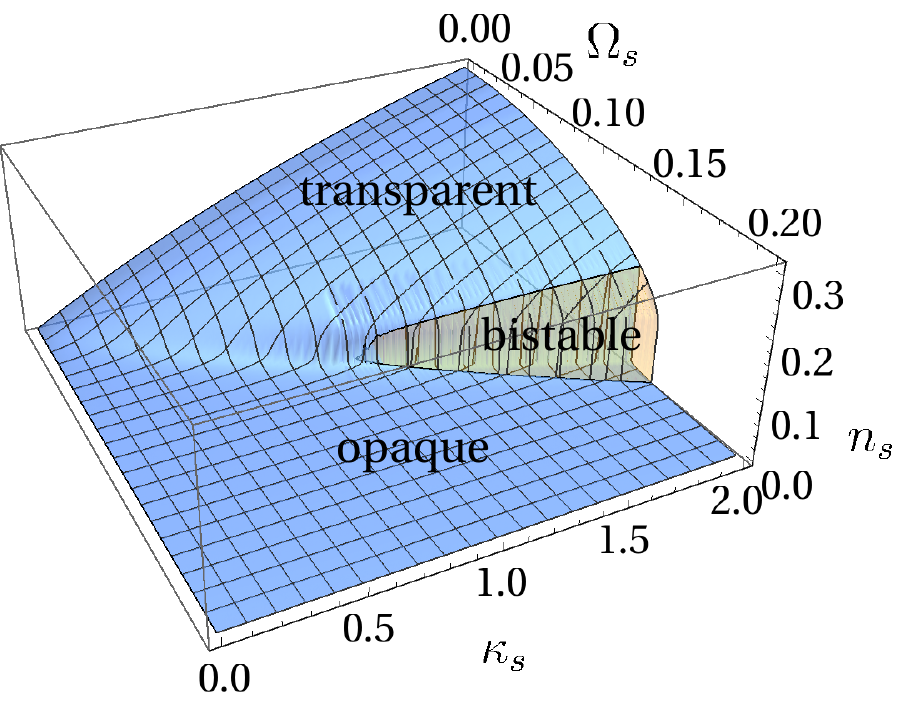}
\caption{Excitation density in the atomic state $s$.
The yellow(blue) surface corresponds to a system
  initialized in the ``transparent"(``opaque") phase with vanishing(large) values of
 $\kappa_s$.  Parameters are $\kappa_0=2$, $\kappa_s=1$, $\omega_0=\Delta_s\equiv\omega_s+\omega_L^{(1)}-\omega_e=\Delta_d\equiv\omega_d-\omega_s-\omega_L^{(2)}=n_V=0$, $g_P=g_E=10$, $\kappa_P=0.5$, $\gamma_d=1$, $\kappa_E=5$, $\Omega=0.2$, $\omega_E(k)=\omega_L^{(2)}-100 k^2$ and $\omega_P(k)=-50 \cos{k}-\omega_e$ in units where $\gamma_e=1$.}
\label{fig:phase_diag}
\end{figure}
In the steady-state the system is translation invariant in both time and
space. This allows to write the retarded M-photon propagator in the most
general form (within our SCH approach) as \footnote{See Supplemental Material}
$
G_M^R(\omega,k)=(\omega-\omega_M(k)-\Sigma_M^R(\omega,k)+i\kappa_M/2)^{-1}
$
with the self-energies of Fig.~\ref{fig:N_scheme_diagrams} given by
\begin{align}
\label{eq:SigmaMsym}
\Sigma_M^R(\omega,k)&=\frac{g_M^2|u_k^M(0)|^2(1-n_V) }{\omega-\omega_e-\frac{\Omega^2}{\omega-\omega_s-\omega_L^{(1)}-\Sigma_s^R\left(\omega-\omega_L^{(1)}\right)}+i\gamma_e/2}\nonumber\\
\Sigma_s^R(\omega)&=\frac{(\Omega_s^\text{eff})^2}{\omega-\omega_d+\omega_L^{(2)}+i\gamma_d/2}
\end{align}
where $n_V$ is the average number of defect-atoms with respect to unit
filling of the photonic crystal.
The standard non-interacting EIT corresponds to setting $\Omega_s^\text{eff}=0$
in Eq.~\eqref{eq:SigmaMsym}. The (imaginary) poles of $G_M^R$ yield
all polariton branches. The effect of the E-photon
mediated interactions in the steady state within our SCH approach is parametrized by the single
real constant $\Omega_s^\text{eff}=\Omega_s|1+\chi|$. The latter satisfies however a non-trivial
integral equation given in the Supplemental Material and represented
diagrammatically in Fig.~\ref{fig:N_scheme_diagrams}.

The phase diagram as a function of the M-photon pump $\kappa_s$ and $s-d$ drive $\Omega_s$ is shown in Fig.~\ref{fig:phase_diag}.
We find two possible steady-state phases: i) an ``opaque''
phase characterized by a small atomic excitation density $n_s$ and ii) a
``transparent'' phase exhibiting instead a much larger $n_s$.
Those two phases are separated by a first order phase transition that
includes a bistable region and terminates in a bi-critical point where
the transition is continuous.
\begin{figure}[t]
\centering
\includegraphics[width=\columnwidth]{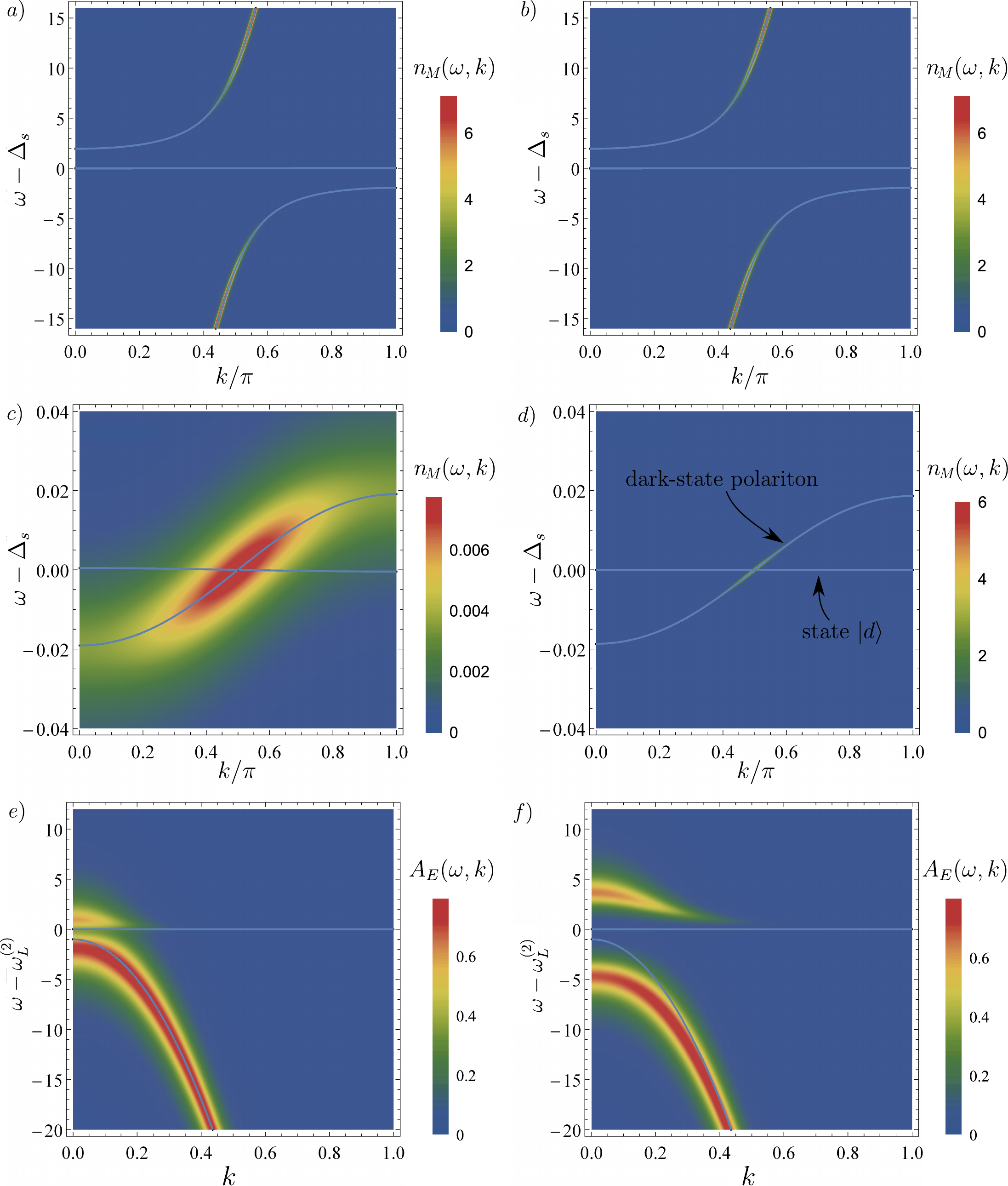}
\caption{Comparison between the opaque (left) and transparent (right)
  solution. Top row: M-photon occupation in
  frequency-momentum space. Solid blue lines correspond the to dispersion curve of
  the polariton branches. Mid-row: enlarged view of the dark-state
  polariton branch. Bottom row: spectral function of the
  E-photons. Solid lines correspond to the bare E-photon
  dispersion (inverted parabola) and the resonance frequency of the $s-d$ transition (horizontal
  line). The
  parameters are the same as in Fig.~\ref{fig:phase_diag} at the bistable point $\Omega_s=0.14$ and $\kappa_s=2$.}
\label{fig:freqmom_res_corr}
\end{figure}

More insight into the properties of the opaque and transparent phases are
obtained by examining the frequency- and momentum-resolved occupation
$n_M(\omega,k)$ shown in Fig.~\ref{fig:freqmom_res_corr}, defined by
$\int_{k,\omega} n_M(\omega,k)=n_M$. In addition to the three polariton
branches of the non-interacting EIT, a fourth branch emerges as a
consequence of the coupling to the level $d$ (see
Fig.~\ref{fig:setup}). Since the coupling is weak it is an almost non-dispersive flat band at $\omega\approx\omega_d+\omega_L^{(1)}-\omega_L^{(2)}$. 

In the opaque phase, the intensity
is almost exclusively distributed on the two outer branches, which are far-off
resonant with respect to the atomic transitions and therefore are not
influenced by the atomic medium. 
The two central branches are essentially empty i.e. no sign of the EIT
window on the dark-state polariton branch (upper branch within the
central pair in Fig.~\ref{fig:freqmom_res_corr}) is left.
The latter is destroyed by coupling the metastable state
$|s\rangle$ to the excited state $|d\rangle$, introducing an
additional decay channel that is eventually inherited by the dark-state polariton.
Correspondingly, the additional polariton branch resulting
from the coupling to the $d$-level hybridizes with the
dark-state polariton.

In the transparent phase on the other hand the intensity is concentrated within a very sharp region around a specific
wavenumber $k_{\rm EIT}$ of the dark-state polariton branch. This
means that in the phase transition the system has reconstructed the transparency window. In the original non-interacting
EIT effect, the window is formed due to the destructive interference
between the two excitation pathways corresponding to the two diagrams for
$\Sigma_M$ in Fig.~\ref{fig:N_scheme_diagrams}. In the IIT effect, the
window is also reconstructed via destructive interference,
this time between the four different excitation pathways involving the
state $|d\rangle$ and corresponding to the first four diagrams
contributing to $\Sigma_s$ in Fig.~\ref{fig:N_scheme_diagrams}.
Other than in the non-interacting EIT, the interfering pathways involve the
E-photons i.e. interactions between polaritons, 
which renders IIT intrinsically
nonlinear. In the lossy system this implies that IIT takes place through a first-order phase transition showing bistability.
The transparency window is reconstructed at a slightly different
position with respect to the non-interacting case with
the EIT-wavenumber still being very well approximated by
$\omega_M(k_{\rm EIT})=\omega_{s}+\omega_L^{(1)}-\omega_g$. 
The destructive interference between the four pathways is most
efficient if the self-energy of the E-photons
$\Sigma_E^R(\omega_L^{(2)}+\omega_s)$ becomes purely imaginary,
corresponding to strong screening between E-photons and the external
laser. This indicates that IIT
is a dissipative many-body effect only accessible to systems far from thermal equilibrium.  

The destruction of the transparency window in the opaque phase via coupling to a
lossy state is an effect analogous to the one employed to build an optical
switch in \cite{bajcsy_2009}, whereby any two-photon
state becomes strongly suppressed. If one uses a lossy state to induce interactions between
atoms in the metastable state, the IIT additionally enables to reconstruct transparency
at a tunable photon number.

The E-photons mediating the interactions between the atoms also
show drastic differences between the opaque and transparent phase. As opposed
to the M-photons, $E$-photons are not driven and can only be excited by
atoms in the $d$ level. The latter can be occupied only via
laser transitions from the $s$ level, which in turn can be populated
via absorption of $M$-photons. Therefore, the occupation of the
electrically polarized mode is suppressed by $1/L_M$ and thus typically small for realistic parameters. It is therefore
more instructive to analyze the spectral function, defined as
$A_E(\omega,k)=-2\Im{G_E^R(\omega,k)}$, which is normalized to $1$ and
accessible for instance by combining the waveguide output with a reference field on
a beam-splitter \cite{buchhold_dis_2013}. $A_E(\omega,k)$ is
shown in the bottom row of Fig.~\ref{fig:freqmom_res_corr}. In the
opaque phase the spectral weight is mostly on the bare dispersion curve,
with a width set by the losses $\kappa_E$. This is caused by the low
number of atoms in state $s$, which makes the modification $\Sigma_E$
of the photon-propagation due to the medium negligible. On the other
hand, in the transparent phase we see that the bare E-photon branch
hybridizes with the atomic $s-d$ excitation. In addition, in the
momentum region close to the dispersion maximum, spectral weight is
transferred from the photon-like branch to the atom-like branch. This
screening effect is quantitatively important and reduces the strength
of the E-photon-mediated interactions for polaritons in the transparency
window. Adiabatic elimination of the bare E-photons would therefore largely overestimate the bright region in the phase diagram.

\emph{Realization.---}
According to the condition
\eqref{eq:strong_coupling}, the emergence of IIT requires strong
atom-light interactions. Fortunately, this requirement is met
for parameters that are expected to become experimentally viable in
PCWs in the near future \cite{kimble_2014_crystal}, namely
$\gamma_{e,d}\sim10\text{MHz}$, $g_{M,E}\sim10^3\gamma_e$,
$J_M\sim10^7\gamma_e$, $\alpha_E\sim10^6\gamma_e$ and
$\kappa_{M,E}\sim10\gamma_e$, for which
our approach indeed predicts the existence of a IIT
  transition. Because of the highly tunable photon dispersions in
PCWs this will likely be possible with $C_E\lesssim 1$, where our
theory becomes quantitatively valid. 
It is also worth mentioning that the additional diagrams that need to
  be added to the ones in Fig.~\ref{fig:N_scheme_diagrams} in order to
render our predictions fully quantitative actually enhance the IIT-effect,
which results in a parametrically larger bistable region
\cite{lang_EIT_long}.


As for the above realistic parameters the relative width of the EIT
window becomes extremely small, in
Figs.~\ref{fig:phase_diag},\ref{fig:freqmom_res_corr} we have, for
illustrative purposes, chosen a different set that however still satisfies $J_M>g_{M,E}>\kappa_{M,E}>\gamma_{e,d}$.

\emph{Conclusions.---}
We introduced the phenomenon of interaction-induced transparency
(IIT), which is characterized by the appearance of a transparency window for
strongly interacting polaritons due to nonlinear interference
effects. In the context of nonlinear quantum optics, IIT constitutes a
novel, genuine quantum many-body effect. From the more fundamental
perspective of many-body physics, the IIT phenomenon is a
non-equilibrium phase transition in the driven-dissipative steady
state which has no analogue so far in condensed matter, as it stems
from the dissipative and retarded nature of the interactions between
polaritons. It would be also interesting to determine whether IIT
belongs to the recently proposed scenario of dark-state phase
transitions \cite{roscher2018phenomenology}.

Future directions could also involve the study of
the role played by IIT in pump-probe experiments, where a light pulse
is locally injected and its transient dynamics is
investigated. 


\acknowledgements
We are grateful to Wilhelm Zwerger for careful reading of the manuscript. DC acknowledges support from ERC Starting Grant FOQAL, MINECO Plan Nacional Grant CANS, MINECO Severo Ochoa Grant No. SEV-2015-0522, CERCA Programme/Generalitat de Catalunya, and Fundacio Privada Cellex.

\bibliography{EIT_biblio}
\newpage
\bigskip
\pagebreak
\widetext
\section{Supplemental Material}

\subsection{The mathematical structure of the SCH approach}

In the diagrammatic representation of our SCH approach, every expression in
Fig.~\ref{fig:N_scheme_diagrams}, corresponds to a propagator (or
Green's function), which in the Keldysh framework can be either the
retarded $G^R$, advanced $G^A$ or the Keldysh $G^K$ propagator. However retarded and advanced Green's functions are not independent and thus, in general
every diagram can generate two self-energy contributions
$\Sigma^{R,K}$. In turn, the internal lines can either be $G^R$ or
$G^K$, provided the contraction is compatible with the interaction
vertex and causality is guaranteed. We skip here all the details of
the derivation of the expressions corresponding to the diagrams in
Fig.~\ref{fig:N_scheme_diagrams}, which are given in \cite{lang_EIT_long}.
We also specify to the translation invariant case discussed in the
main text, where the photon propagators depend on frequency and only a
single momentum, while the atom propagators depend solely on the frequency.
The $M$-photon propagator reads:
\begin{align}
G_M^R(\omega,k)&=\left[G_M^A(\omega)\right]^*=\frac{1}{\omega-\omega_M(k)-\Sigma_M^R(\omega,k)+i\kappa_M/2}\nonumber\\
G_M^K(\omega,k)&=G_M^R(\omega,k)\left(\Sigma_M^K(\omega,k)-i\kappa_M-2i\kappa_s\right)G_M^A(\omega,k),
\label{eq:GMsym}
\end{align}
with
\begin{align}
\label{eq:SigmaMsym}
\Sigma_M^R(\omega,k)=\frac{g_M^2|u_k^M(0)|^2(1-n_V) }{\omega-\omega_e-\Omega^2 \tilde{G}_s^R(\omega+\omega_L^{(1)})+i\gamma_e/2}
\nonumber\\
\Sigma_M^K(\omega,k)=2i\mathrm{Im}\Sigma_M^R(\omega,k)
\end{align}
where $1-n_V$ is the atom filling factor in the photonic crystal
waveguide
and the effective $s$-atom propagator
\begin{align}\label{eq:GsR_sym}
\tilde{G}_s^R(\omega)&=\left[\tilde{G}_s^A(\omega)\right]^*=\frac{1}{\omega-\omega_s-\tilde{\Sigma}_s^R(\omega)+i0⁺/2}
\end{align}
with
\begin{align}
\tilde{\Sigma}_s^R(\omega)&=\frac{\left(\Omega_s^\text{eff}\right)^2}{\omega+\omega_L^{(2)}-\omega_d+i\gamma_d/2}\,.
\end{align}
Here $\Omega_s^\text{eff}=\Omega_s^2\left(1+\chi\right)^2$ is the effective Rabi amplitude modified by the $E$ photons, which we parametrize by the dimensionless, complex constant $\chi$.
Note that this construction of $\Sigma_M^R$ takes into account the fact that we have to exclude all diagrams
where the same atoms appear in the same state twice. This is achieved
by removing the last diagram for $\Sigma_s$ in
Fig.~\ref{fig:N_scheme_diagrams} whenever $s$ atoms appear inside the
polarization loop of the $M$-photons.
For the same reason, in the first four diagrams for $\Sigma_s$, the
internal $s$-propagators are allowed to contain only the last diagram
as self-energy insertion. This defines a second type of $s$-propagator:
\begin{align}\label{eq:GsR_sym}
\tilde{\tilde{G}}_s^R(\omega)&=\left[\tilde{\tilde{G}}_s^A(\omega)\right]^*=\frac{1}{\omega-\omega_s-\tilde{\tilde{\Sigma}}_s^R(\omega)+i0^+/2}
\end{align}
with
\begin{align}
\tilde{\tilde{\Sigma}}_s^R(\omega)&=\frac{\Omega^2}{\omega-\omega_e+\omega_L^{(1)}-
  \Sigma_{e}^R\left(\omega+\omega_L^{(1)}\right)+i\gamma_e/2}
\end{align}
and
\begin{align}
\tilde{\tilde{\Sigma}}_s^K(\omega)&=2i\mathrm{Im}\tilde{\tilde{\Sigma}}_s^R(\omega)+\delta \tilde{\tilde{\Sigma}}_s^K(\omega)\nonumber\\
\delta \tilde{\tilde{\Sigma}}_s^K(\omega)&=\frac{\Omega^2\left(\Sigma^K_{e} \left(\omega+\omega_L^{(1)}\right)-2i\mathrm{Im}\Sigma^R_{e} \left(\omega+\omega_L^{(1)}\right)\right)}{\left(\omega-\omega_e+\omega_L^{(1)}-\mathrm{Re}\Sigma^R_{e} \left(\omega+\omega_L^{(1)}\right)\right)^2+\left(\gamma_e/2-\mathrm{Im}\Sigma^R_{e}
\left(\omega+\omega_L^{(1)}\right)\right)^2},
\end{align}
where
\begin{align}
\Sigma^R_{e}(\omega)&=-\sum_n \int\frac{dk}{2\pi}g_M^2\frac{\kappa_s}{2\mathrm{Im}(\omega_n(k))}\frac{1}{\omega-\omega_n(k)+i0^+/2}\frac{f(\omega_n(k)) f^*(\omega_n^*(k))}{\prod_{m\neq n}(\omega_n(k)-\omega_m(k)) (\omega_n(k)-\omega_m^*(k))} \left|u_k^M(0)\right|^2 \notag \\
&+\int \frac{dk}{2\pi} \frac{1}{2}g_M^2(4-2n_V)|u_k^M(0)|^2 G_M^R(\omega+i0^+/2,k)\nonumber\\
\Sigma^K_{e}(\omega)&=2i\mathrm{Im}\Sigma^R_{e}(\omega)-i\kappa_s \int \frac{dk}{2\pi} g_M^2(2-2n_V)|u_k^M(0)|^2 G_M^R(\omega,k)G_M^A(\omega,k),
\end{align}
where $n \in \{1,2,3,4\}$, $\omega_n(k)$ are the poles of $G_M^K(\omega,k)$ and
\begin{align}
f(\omega)=&(\omega-\omega_e+i\gamma_e/2)(\omega-\omega_s-\omega_L^{(1)}+i0^+/2)(\omega-\omega_d-\omega_L^{(1)}+\omega_L^{(2)}+i\gamma_d/2) \notag \\&- \Omega^2(\omega-\omega_d-\omega_L^{(1)}+\omega_L^{(2)}+i\gamma_d/2)-(\Omega_s^\text{eff})^2(\omega-\omega_e+i\gamma_e/2)\;.
\end{align}

The constant $\chi$ in the effective Rabi amplitude $\Omega_s^\text{eff}=\Omega_s|1+\chi|$ is then to be determined self-consistently. The
corresponding equation is
\begin{align}\label{eq:Cconst}
\chi=\frac{\Sigma_E^R(\omega_L^{(2)})}{\omega_L^{(2)}-\omega_E(0)-\Sigma_E^R(\omega_L^{(2)})+i\kappa_E/2},
\end{align}
where the $E$-photon self-energy reads
\begin{align}
\Sigma_E^R(\omega)=\frac{i}{2}\int\frac{d\omega'}{2\pi}g_E^2 \tilde{\tilde{G}}_s^R(\omega') \tilde{\tilde{G}}_s^A(\omega')\delta \tilde{\tilde{\Sigma}}_s^K(\omega')\frac{1}{\omega'+\omega-\omega_d+i\gamma_d/2}.
\end{align}
The latter describes the modification of the $E$-photon propagation,
and thus of the light-mediated atom-atom interactions, due to the
medium. The photon propagator is then given by
\begin{align}
G_E^R(\omega,k)&=\left[G_E^A(\omega)\right]^*=\frac{1}{\omega-\omega_E(k)-\Sigma_E^R(\omega,k)+i\kappa_E/2}\nonumber\\
G_E^K(\omega,k)&=G_E^R(\omega,k)\left(\Sigma_E^K(\omega,k)-i\kappa_E\right)G_E^A(\omega,k),
\label{eq:GMsym}
\end{align}
with 
\begin{align}
\Sigma_E^K(\omega,k)=2i\mathrm{Im}\Sigma_E^R(\omega,k).
\end{align}

\end{document}